\documentstyle[aps,epsf]{revtex}
\begin{document}

\begin{flushright}
JLAB-THY-98-45 \\
August 1998
\end{flushright}

\vspace{2cm}

\begin{center}
{\Large \bf Compton Scattering  and 
Nonforward 
Parton Distributions\footnotemark}
\end{center}

\footnotetext{Contribution to Proceedings 
of the Workshop ``Physics and Instrumentation
with 6-12 GeV  Beams'', Jefferson Lab, June 15-18, 1998}

\begin{center}
{A.V. RADYUSHKIN\footnotemark}  \\
{\em Physics Department, Old Dominion University,}
\\{\em Norfolk, VA 23529, USA}
 \\ {\em and} \\
{\em Jefferson Lab,} \\
 {\em Newport News,VA 23606, USA}
\end{center}

\footnotetext{Also  Laboratory of Theoretical Physics,
JINR, Dubna, Russian Federation}

\newpage

\noindent {\bf Compton amplitudes.}
Compton scattering  (see Fig.1) in its various versions 
provides a unique tool for studying 
many aspects of hadronic structure  probing it  by 
two electromagnetic currents. 
In QCD, the photons couple to 
quarks of  a hadron  through a   
vertex  which is pointlike in the simplest
 approximation. 
However, in the soft regime, 
strong interactions  produce large 
corrections uncalculable within the
perturbative QCD framework.
To  take advantage of the basic  pointlike structure of the
photon-quark coupling, one should choose 
a specific kinematics  securing the short-distance dominance 
by the 
presence of a large momentum transfer.

\begin{figure}[t]
\mbox{
   \epsfxsize=11cm
 \epsfysize=5cm
  \epsffile{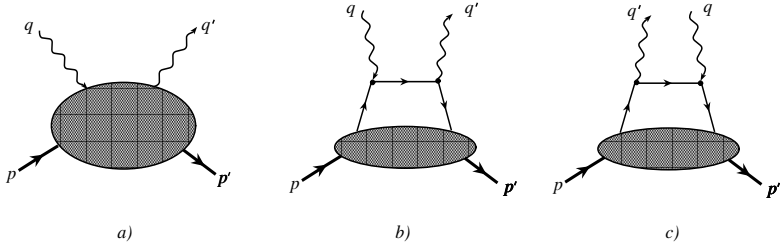}  }
\vspace{-1.5cm}
{\caption{\label{fig:cohan} $a)$ General Compton amplitude;
$b)$ $s$-channel handbag diagram;  $c)$ $u$-channel handbag diagram.
   }}
\end{figure}

{\it Deep inelastic scattering.}
Virtual forward Compton amplitude whose
imaginary part gives structure functions
of deep inelastic scattering is the most
well-known case of a 
short-distance-dominated Compton amplitude.
 In this case,
the ``final'' photon  has momentum $q'=q$ coinciding with
that of the initial one. The momenta $p, p'$ of the initial
and final hadrons also coincide. 
The short-distance dominance is guaranteed by high
virtuality of the photons: $-q^2 \equiv Q^2 > 1 $ GeV$^2$.
Moreover, the total cm energy of the photon-hadron system
$s= (p+q)^2$ should be above resonance region,
and the Bjorken ratio $x_{Bj} = Q^2/2(pq)$ is finite.
In the large-$Q^2$ limit,  the dominant contribution 
is given by the handbag diagram in which the 
perturbatively calculable hard quark
propagator is convoluted with parton distribution
functions $f_a(x)$ ($a=u,d,s, \ldots$) 
which describe/parametrize nonperturbative
information about hadronic structure.

{\it Deeply virtual Compton scattering (DVCS).}
In this case, the initial photon is highly
virtual while the final photon is real.
The momentum transfer $t$ should be 
as small as possible \cite{ji}. 
 Large  virtuality 
of the initial photon is sufficient for 
making the handbag diagram dominant \cite{npd,jo,cf}.  
 Again, the leading term  is given by a convolution
of the hard quark propagator and 
a nonperturbative function ${\cal F}_{\zeta}(X;t)$
({\it nonforward parton distribution}\footnote{X.Ji introduced 
originally {\it off-forward parton distributions} $H(x,\xi;t)$ \cite{ji}, with 
parton momentum being $xP$, where $P=(p+p')/2$. } \cite{npd} ) 
describing  long-distance dynamics.
In addition to the ``usual'' parton variable $X$
giving  the fraction $Xp^+$ of the  ``$+$'' component
($p^+ = p^0+ p^3$) of the initial 
hadron momentum $p$ carried by the quark, 
it also depends on 
the invariant momentum transfer $t=(p'-p)^2$,
and the ``skewedness'' parameter $\zeta = r^+/p^+$ 
(where $r \equiv p-p'$)  specifying the longitudinal
momentum asymmetry of the nonforward matrix element.
This asymmetry appears because it is impossible 
to convert a highly virtual initial photon into a real
final photon without a longitudinal momentum transfer
(the longitudinal direction is specified 
by $p$ and $q$).  For the DVCS amplitude,
the skewedness parameter $\zeta$ coincides with 
the Bjorken variable $x_{Bj}$\footnote{In Ji's approach \cite{ji}, the
skewedness  is characterized by the parameter $\xi = x_{Bj}/(2-  x_{Bj}) $.} .

{ \it  Wide-angle real Compton scattering (WACS).}
With both photons real, it is not sufficient to have
large photon energy to ensure short-distance dominance:
large-$s$, small-$t$ region is dominated by
Regge contributions for which no simple
parton description is possible.
Hence, having large $|t| > 1$GeV$^2$ is a  necessary
condition for revealing short-distance  dynamics.
The leading-order contribution in the 
large-$s$, large-$t$ limit is again given by
handbag diagrams \cite{realco} in which  hard quark propagator
is convoluted with the nonforward parton distributions
 $ {\cal F}^a_{\zeta=0}(X;t)$. 
Since the initial photon is real,
it can be converted into the final one by a purely
transverse momentum transfer, so the skewedness parameter
$\zeta$ vanishes. 

\noindent  { \bf Hybrid nature of nonforward parton distributions (NFPD's).}
The
NFPD's   $ {\cal F}^a_{\zeta}(X;t)$
look like form factors with respect to $t$ 
and like parton distributions with respect to
$X$.  Due to  the spectral 
property  $0 \leq X \leq 1$,
both the parton going out of the hadron 
and the spectators carry positive fractions 
of the initial hadron momentum
(parton picture implies a frame where the hadron moves fast).

\begin{figure}[ht]
\mbox{
   \epsfxsize=11cm
 \epsfysize=5cm
\hspace{1cm} 
  \epsffile{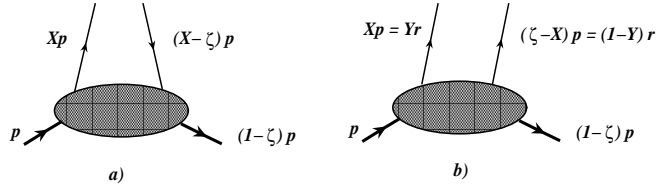}  } 
\vspace{-2.3cm}
{\caption{\label{fig:nonfwd} Nonforward parton  distribution
in  two regions:  $a)$ $X > \zeta$ and 
$b)$  $X < \zeta$. 
   }}
\end{figure}

Unlike in the forward case, 
the ``returning''
parton carries   a smaller fraction $X' \equiv X- \zeta$    
of the original hadron momentum $p$.   
Since $0 < \zeta <1$, the fraction 
$X'$ can be either positive  or negative, ${\rm i.e.},$  
 each NFPD  has two components. 
In the region $\zeta< X < 1$ (Fig. \ref{fig:nonfwd}$a$),
the function ${\cal F}_{\zeta} (X)$ 
 describes
 a parton   going out of 
the hadron with a positive fraction  $Xp$  
of the original hadron momentum
and then coming back into the hadron with a changed 
(but still positive) fraction  $(X - \zeta)p$. 
In the region  $0< X < \zeta$ (Fig.  \ref{fig:nonfwd}$b$), 
 the ``returning''  parton   has
a negative fraction $(X- \zeta)$ of the light-cone momentum $p$.
Hence, it is more appropriate to  treat it  as a parton 
going out of the hadron and 
propagating  along  with the original parton.
Writing $X$ as $X = Y \zeta$, we see that 
both   partons  carry now 
positive fractions $Y \zeta p \equiv Y r$ and
  $(1-Y)\, r $ of 
the momentum transfer $r$.  In the region $X= Y \zeta < \zeta$,
NFPD's 
look like  distribution amplitudes  (wave functions) 
$
\Psi_{\zeta}(Y) =   {\cal F}_{\zeta}(\zeta Y)$
characterizing  the
probability amplitude for the initial hadron with momentum 
$p$ to split into the final hadron with momentum $(1-\zeta)p$
and a two-parton state with total momentum $r=\zeta p$
shared by the partons
in fractions $Yr$ and $(1-Y)r$
(see Fig.\ref{fig:nonfwd}).

\begin{figure}[htb]
\mbox{
   \epsfxsize=7cm
 \epsfysize=4cm
 \hspace{2cm}  
  \epsffile{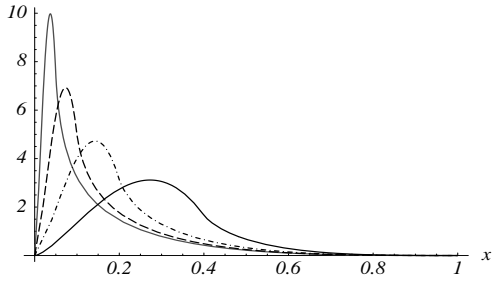}  }
  \vspace{0.5cm}
{\caption{\label{fig:3} Nonforward parton distributions 
$ {\cal F}_{\zeta}(X)$ for different values of the skewedness
$\zeta=0.05$ (thin line), $\zeta=0.1$ (dashed line),
$\zeta=0.2$ (dash-dotted line) and $\zeta=0.4$ (full line)
in the ``valence quark oriented''  model of ref.  [6].
   }}
\end{figure}

Possible shape  of nonforward distributions
$ {\cal F}_{\zeta}(X)$  is  illustrated in Fig.\ref{fig:3}.
A characteristic feature of each curve 
is a  maximum located close to 
 the relevant border 
 point $X = \zeta$ and slightly shifted 
 to the left from it\footnote{More exotic results 
with rapid variation of  NFPD's in the region $X \sim \zeta$ 
were obtained in the chiral soliton model \cite{max}.}.
The curves shown there were obtained in a model \cite{ddee} 
based on so-called double distributions 
(DD's)   $F(x,y;t)$  which 
provide an alternative  parametrization \cite{npd}
of nonforward matrix elements, 
in which the parton momentum is 
written as $xp+yr$. DD's behave like parton densities 
wrt $x$ and like distribution amplitudes wrt $y$.

Formally, 
nonforward parton distributions are definined through 
matrix elements of  light-cone operators
\begin{eqnarray} 
&&\lefteqn{
\langle \, p'  , s' \,  | \,   \bar \psi_a(0)  \hat z 
E(0,z;A)  \psi_a(z) 
\, | \,p ,  s \,  \rangle |_{z^2=0}
 } \label{57}  \\ 
&&  
= \bar u(p',  s')  \hat z 
 u(p,s) \, \int_0^1   
 \left ( e^{-iX(pz)}
 {\cal F}^a_{\zeta}(X;t)  -  e^{i(X-\zeta)(pz)}
{\cal F}^{\bar a}_{\zeta}(X;t)  \right ) 
 \,  dX \nonumber \\ 
&&  
+
\, \bar u(p',s')  \frac {\hat z  \hat r - \hat r \hat z}{4M}
 u(p,s) \, \int_0^1   
  \left ( e^{-iX(pz)}
 {\cal K}^a_{\zeta}(X;t)  -  e^{i(X-\zeta)(pz)}
{\cal K}^{\bar a}_{\zeta}(X;t)  \right ) 
 \,  dX
.
\nonumber
 \end{eqnarray} 
Here $E(0,z;A) $ is the straight-line gauge link, $M$ is the nucleon mass, 
$s,s'$ specify the nucleon  polarization and  $\bar u(p',  s') , 
 u(p,s) $  are Dirac spinors. We use the
``hat''  convention $\hat z \equiv z^{\mu} 
\gamma_{\mu}$.

Using this definition, one can relate   NFPD's 
to simpler functions which  have been already  studied
experimentally.
For instance, if $p=p'$, 
the matrix element coincides with  the 
forward one defining  usual  parton distribution functions.
This  results in 
$$ {\cal F}^a_{\zeta=0}(X,t=0) = f_a(X) \, ;  \, 
{\cal F}^g_{\zeta=0}(X,t=0) = Xf_g(X).
$$
Taking $z \to 0$, we see that integrating ${\cal F}_{\zeta}(X,t)$ over $X$
one obtains hadronic form factors: 
$$\sum_a e_a \int_0^1   
 \left [
 {\cal F}^a_{\zeta}(X;t)  -  
{\cal F}^{\bar a}_{\zeta}(X;t)  \right ]
 \,  dX  =F_1(t) \, , $$
where $e_a$ is the electric charge of the
relevant quark.   The second type of distributions
$ {\cal K}^{ a}_{\zeta}(X;t) $  present in Eq.(\ref{57}) 
correspond to hadron helicity flip in the 
nonforward matrix element. They 
are related to the $F_2(t)$ form factor. 
Note, that  the ${\cal K}$ functions are accompanied 
by the $r$ factor in the above parametrization
of the nonforward matrix element.
Hence,  they are invisible
in deep inelastic scattering  and other inclusive processes 
described 
by  exactly  forward $r=0$ matrix elements.  
However, the $t=0, \zeta =0 $ limit of the ${\cal K}$ distributions 
${\cal K}^a_{\zeta=0}(X;t=0) \equiv k_a(X)$ exists.
In particular, the integral
$$
\sum_a e_a \int_0^1   
 \left [ k_a(X) - k_{\bar a}(X) 
  \right ] 
 \,  dX  = \kappa_p \, 
$$
gives the proton anomalous magnetic moment. 
The $X$-moment of the flavor-singlet combination 
$k_a(X) + k_{\bar a}(X) $ 
contributes to the orbital momentum contribution
to the proton spin \cite{ji}. 
Furthermore, there are also parton-helicity 
sensitive nonforward distributions 
$ {\cal G}^{ a}_{\zeta} (X;t)$ and  
$ {\cal P}^{ a}_{\zeta} (X;t)$. 
The first one  
reduces to the usual spin-dependent 
densities $\Delta f_a(x)$ in the 
$r=0$ limit and gives the axial 
form factor $F_A(t)$ after the $X$-integration.
The second one is  similarly related to the pseudoscalar form factor 
 $F_P(t)$.

\noindent { \bf Deeply virtual Compton scattering.}
In the lowest order,  the DVCS amplitude 
$T^{\mu \nu} (p,q,q')$  is given by two
handbag diagrams.  In particular, the invariant  amplitude 
containing the ${\cal F}$ functions is given by
\begin{eqnarray} 
T_F(p,q,q') = \sum_a e_a^2 
 \int_0^{1} 
\left  [ \frac1{X-\zeta +i\epsilon}
+ \frac1{X- i \epsilon} \right ]  {\cal F}^{a+ \bar a}_{\zeta}(X;t)
\, dX \,  . 
\label{123}  \end{eqnarray} 
 An important  feature of the DVCS amplitude is that 
for large $Q^2$ and fixed $t$ it depends only on the ratio 
$Q^2/2 (pq) \equiv x_{Bj}
=\zeta$:  DVCS is an  {\it exclusive} process 
exhibiting the Bjorken  scaling. 

One may ask which  $Q^2$   are large enough to ensure 
the dominance of the lowest-twist  handbag contribution. 
In DIS, approximate Bjorken scaling  starts at $Q^2 \sim 2 \, $GeV$^2$.
Another example  is given by the exclusive
process  $\gamma(q_1) \gamma^*(q_2) \to \pi^0$ studied on 
$e^+ e^-$ colliders.  If one of the photons  is highly virtual
$q_1^2 = -Q^2$ while another is (almost) real $q_2^2 \sim 0$,
the process is kinematically
similar  to DVCS.  
In the leading order, the $F_{\gamma \gamma^* \pi^0}(Q^2)$ 
transition form factor  is  given by a handbag diagram again. 
The recent measurements by CLEO \cite{cleo} 
show that the pQCD prediction $F_{\gamma \gamma^* \pi^0}(Q^2) \sim 1/Q^2$ 
again works starting  from $Q^2 \sim 2 \, $GeV$^2$.
The $\gamma \gamma^* \pi^0$ vertex  (for a virtual pion) can be also 
measured on a fixed-target machine  like CEBAF 
in which case it is just a part
of the DVCS amplitude  corresponding to the 4th nonforward distribution
${\cal P}_{\zeta}(X,t)$ (which is related to the pseudoscalar 
form factor  $G_P(t)$ of  the nucleon).
Hence, CLEO data  give an evidence that DVCS  may be 
handbag-dominated for $Q^2$ as low as 2\,GeV$^2$.

The main problem for studying DVCS is the contamination 
by the Bethe-Heitler process in which the final photon
is emitted from the initial or final electron.
The Bethe-Heitler amplitude is enhanced at small $t$.
On the other hand, the virtual photon flux 
for fixed $Q^2$ and $x_{Bj}$ increases when 
the electron beam energy increases.  Hence,  the  energy  upgrade 
 would make the DVCS studies  at  Jefferson Lab 
more feasible. Experimental aspects  of virtual
Compton scattering studies at Jefferson Lab were discussed
at this workshop 
in the talk by C.E. Hyde-Wright \cite{charles}.

The nonforward parton distributions can be also 
measured  in hard meson electroproduction processes
\cite{npd,cfs,lech,guivan}.   The leading-twist pQCD contribution in this case
involves a  one-gluon exchange, which means that the hard
subprocess  is suppressed by  $\alpha_s/\pi \sim 0.1$ 
factor. The competing soft mechanism corresponds to
a triple overlap of hadronic wave functions and 
has   a relative  suppression  $M^2/Q^2$ by a power of  
$Q^2$,  with $M^2 \sim 1$\, GeV$^2$  being a characteristic
hadronic scale. Hence, to clearly see the 
one-gluon-exchange signal one needs $Q^2$ above 10\,GeV$^2$.
Numerical pQCD-based estimates 
and comparison of DVCS and hard meson electroprodution 
cross sections \cite{guivan} 
were presented at this workshop  by M. Guidal
\cite{michel}.  

\noindent {\bf Wide-angle real Compton scattering.}
At moderately large $s, |t|, |u| \lesssim \, 10 \,$ GeV$^2$, the  leading-order
contribution  to WACS 
amplitude is given by the $s$- and $u$-channel handbag
diagrams with hard quark propagators 
$(x \hat P + \hat Q)/xs$  and  $(x \hat P - \hat Q)/xs$,
respectively,  where $P=(p+p')/2, Q=(q+q')/2$.
 Since there is no skewedness, one
deals  with the functions  ${\cal F}^a(x,t) \equiv {\cal F}^a_{\zeta=0}(x;t)$
and their ${\cal K,G,P} $ analogs.
The {\it nonforward parton densities}  (ND's) ${\cal F}^a(x,t) $
are 
 the  simplest hybrids of parton densities and hadronic 
form factors.  
They  are 
related  to the usual parton densities by
${\cal F}^a(x,t=0) = f^a(x)$ and to form factors  by 
\begin{equation}
\sum_a e_a \int_0^1   
 \left [
 {\cal F}^a (x;t)  -  
{\cal F}^{\bar a}(x;t)  \right ]
 \,  dX  =F_1(t) \, . \label{nd-ff}
\end{equation}
These relations are trivially satisfied by 
the factorized ansatz ${\cal F}^a_{val} (x;t)  
=  f^a_{val} (x) F_1(t) $ (here ${\cal F}^a_{val} (x;t) 
\equiv  {\cal F}^a (x;t)  -   {\cal F}^{\bar a}(x;t) $). 
However, explicit calculations within the light-cone 
formalism suggest a more complicated interplay 
between the $x$- and $t$-behavior of ND's.
In particular, assuming a relativistic Gaussian dependence 
$\exp[-k^2_{\perp}/2x \bar x \lambda^2]$  ($\bar x \equiv 1-x$) 
of the hadronic blob 
on the transverse momentum $k_{\perp}$,
gives  the model \cite{realco} 
\begin{equation}
{\cal F}^a(x,t) = f^a(x) e^{\bar x t /4 x \lambda^2 }  = 
{{f_a(x)}\over{\pi x \bar x  \lambda^2}}\,
\int  \, e^{-(k^2_{\perp}+ (k_{\perp}+
\bar x r_{\perp})^2)/2x \bar x \lambda^2}
d^2 k_{\perp} \,  . \label{13}
\end{equation}
The functions  $f_a(x)$  here are the 
usual parton densities assumed to be 
taken from existing  parametrizations like GRV, MRS, CTEQ, etc.

\begin{figure}[htb]
\mbox{
   \epsfxsize=5cm
 \epsfysize=5cm
 \hspace{3cm}  
  \epsffile{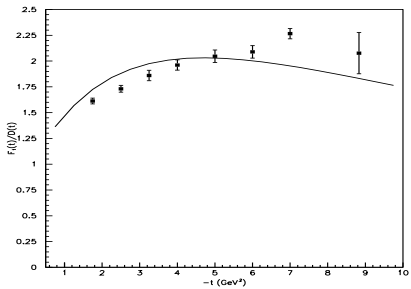}  }
{\caption{\label{fig:ff} Ratio $F_1^p(t)/D(t)$ 
of the $F_1^p(t)$ form factor to
the dipole fit $D(t) =1/(1-t/0.71\,{\rm GeV^2})^2$. Curve
is based on  the model having  a Gaussian 
 $k_{\perp}$-dependence (4) with $\lambda ^2 = 0.7 \,
{\rm GeV}^2$. 
   }} 
\end{figure}

The  basic scale $\lambda$ specifies  the 
  average transverse momentum carried by the quarks. 
The magnitude of  $\lambda$ can be fixed  from 
  the  relation (\ref{nd-ff}) between
  ${\cal F}^a(x,t)$'s
and $F_1(t)$ form factor.
The best agreement  with experimental data \cite{ff} in the
  moderately large $t$ region
   \mbox{1  GeV$^2$ $< |t|< 10$ GeV$^2$}  is reached for 
   $\lambda^2 =0.7 \,$ GeV$^2$ (see Fig.\ref{fig:ff}). 
   This value also gives a reasonable magnitude 
 $ \langle  k^2_{\perp} \rangle^u = (290 \,  {\rm MeV})^2 \, 
  ,    \, \langle  k^2_{\perp} \rangle^d = (250 \, {\rm MeV})^2$ 
   for the average transverse momentum of the valence $u$ and $d$ quarks
   in the proton.   Similar models can be constructed for
$F_2(t)$ and pion form factor $F_{\pi}(t)$ \cite{af}.

It should be emphasized that the large-$t$ 
real Compton scattering amplitude contains 
integrals 
with  an extra $1/x$ factor. 
This results in  a rather  large 
enhancement of the Compton amplitude in the wide-angle regime.
Keeping only the enhanced terms gives 
the WACS  cross section  as a product 
 \begin{equation}
 \frac{ d \sigma}{dt}  \approx \frac{2 \pi \alpha^2}{\tilde s^2} 
 \left [  \frac{(pq)}{(pq')} +  \frac{(pq')}{(pq)}
  \right ] \,  R_1^2(t) \,  , 
 \end{equation}
 of  the  Klein-Nishina  cross section 
and 
 the square of a new  effective form factor $R_1(t)$. In the
model of ref.\cite{realco},  $R_1(t)$ is given by    
 \begin{equation} 
R_1(t) =  \sum_a e_a^2   \int_0^1 
f_a (x) 
  e^{\bar x t / 4x \lambda^2} \frac{dx}{x} \, . 
\end{equation}

\begin{figure}[htb]
\mbox{
   \epsfxsize=5cm
 \epsfysize=7cm
  \epsffile{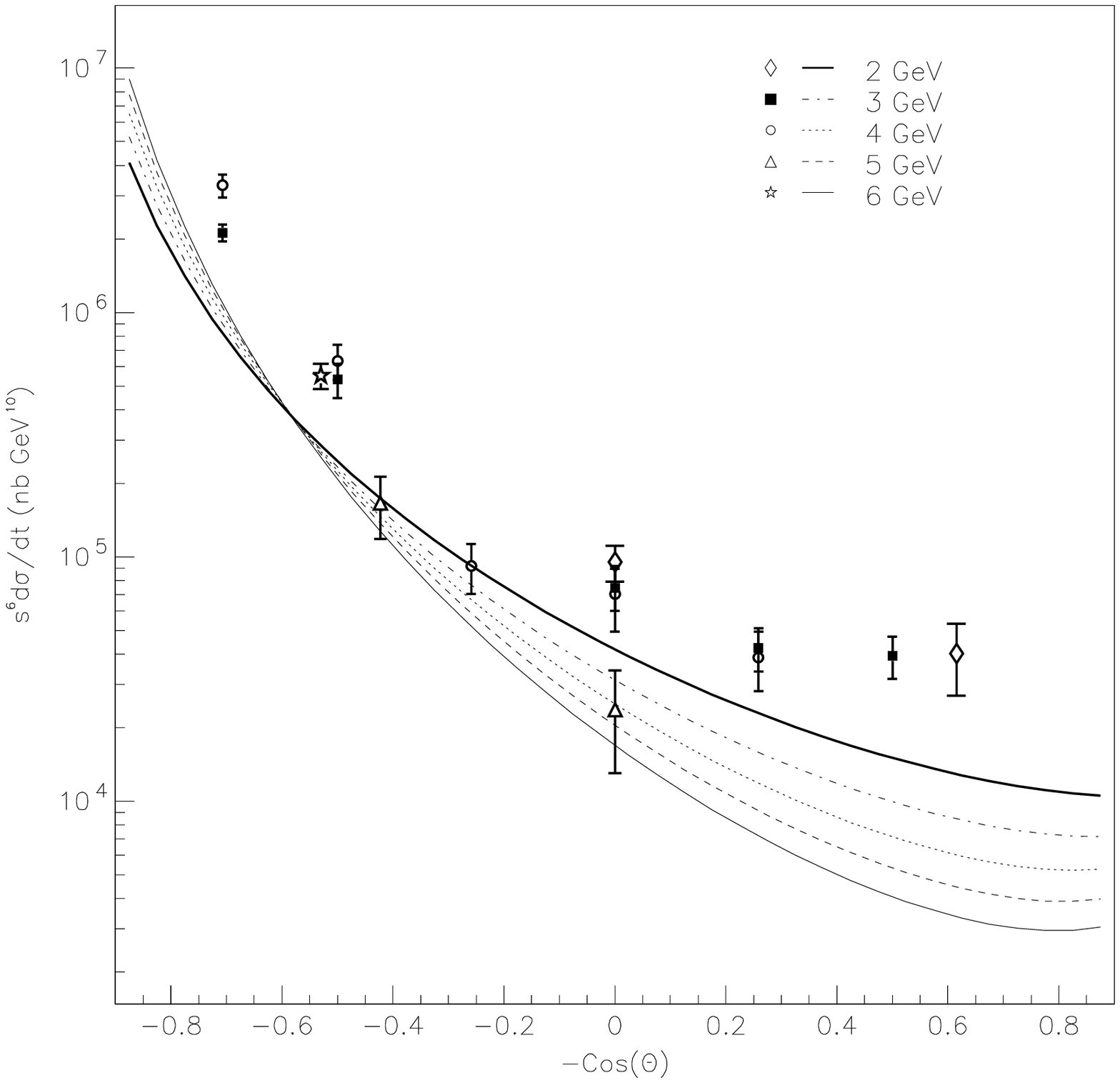}  
 \epsfxsize=5cm
 \epsfysize=7cm
\hspace{0.5cm}  
 \epsffile{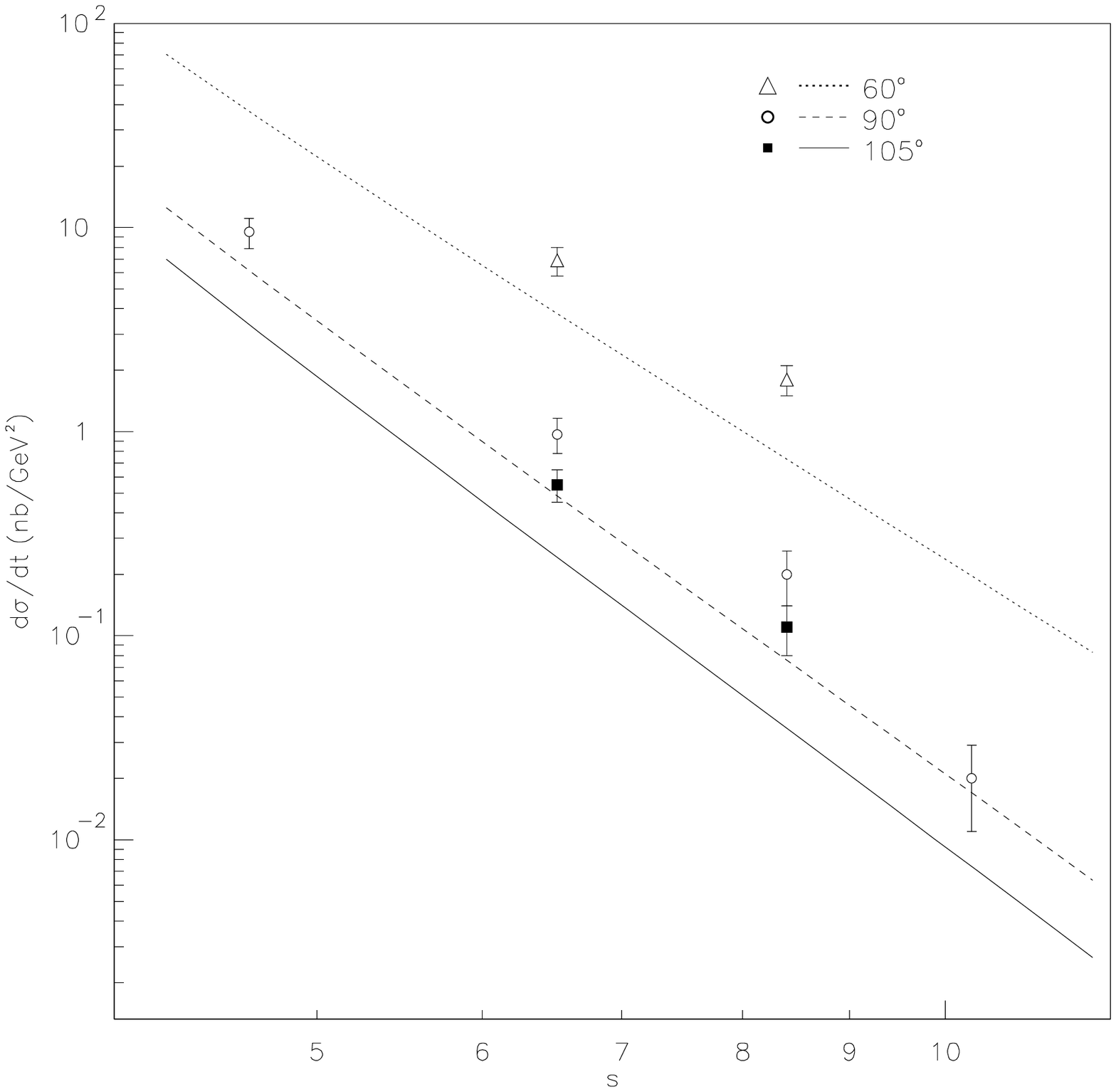} 
}
  \vspace{-0.7cm}
{\caption{\label{fig:rct} $a)$ 
Angular dependence of 
the combination $s^6 (d \sigma /dt)$. $b)$ $s$-dependence of 
the differential cross section $d \sigma /dt $ 
 for $\theta = 60^{\circ}$
(dotted line),  
$\theta = 90^{\circ}$ (dashed line)
and $\theta = 105^{\circ}$ (solid line).
   }}
\end{figure}

Comparison with existing data \cite{schupe} 
is shown in Fig.\ref{fig:rct}.  Our curves  follow the data pattern
but are systematically lower  by a factor of  2. 
Since we neglected several terms each capable 
of producing up to a $20 \%$ correction in  the amplitude, we consider
the agreement between our curves and the data 
as encouraging.

The angular dependence of our results for the combination
$s^6 (d \sigma /dt)$ is shown on Fig.\ref{fig:rct}a.
All the curves for initial photon ehergies 2,3,4,5 and 6 GeV
intersect each other at  $\theta_{\rm cm} \sim 60^{\circ}$.
This  is in good agreement with experimental data
of ref.\cite{schupe} where the differential 
cross section at fixed cm angles was fitted also by powers of $s$:
$d \sigma /dt \sim s^{-n (\theta)}$ with 
$n^{\rm exp}(60^{\circ}) = 5.9 \pm 0.3$. 
Our curves correspond to $n^{\rm soft}(60^{\circ}) \approx 6.1$
and $n^{\rm soft}(90^{\circ}) \approx 6.7$ which also agrees 
with the experimental result $n^{\rm exp}(90^{\circ}) = 7.1 \pm 0.4$.

This can be compared with the scaling behavior
of the asymptotic  hard contribution: 
modulo logarithms contained in the $\alpha_s$ factors,
they have    a universal angle-independent power
$n^{\rm hard}  (\theta) =6$.
For this reason, it is very important
to perform precision measurements of $n(\theta)$ in
future experiments  at Jefferson Lab \cite{alan}.

Above picture for the large-$s$, high-$t$ behavior
of the real Compton amplitude implies  that
the large-$t$ behavior of the form factor type
matrix elements is determined by an overlap
of soft components  of hadronic wave functions.
It should be mentioned that for
extremely large $|t| \gg 10\,$GeV$^2$ 
this mechanism is subdominant.
The dominant contribution is provided
by hard gluon-exchange diagrams.
There are all reasons to expect, however,  that at 
 accessible $t$  the latter 
are negligibly small \cite{realco}.

\noindent 
{\bf Conclusions.}
The hard exclusive electroproduction processes 
provide   new  information about 
hadronic structure accumulated in 
nonforward  parton distributions.  The NFPD's are   
universal  hybrid functions having the 
properties of parton densities,  hadronic form factors 
and distribution
amplitudes.    They give a unified description of  
various hard  exclusive and inclusive reactions. 
The basic supplier of information
about nonforward parton distributions is 
deeply virtual Compton scattering   which 
offers a remarkable 
example of Bjorken scaling 
phenomena  in 
exclusive processes.  
Wide-angle real Compton scattering is an ideal tool
to test angle-dependent scaling laws characteristic 
for soft overlap mechanism.
Hard meson electroproduction is the best candidate 
to see pQCD hard gluon exchange in exclusive reactions.

\end{document}